\documentclass{ws-procs9x6}

\begin{document}

\title{INTEGRAL - OPERATING HIGH-ENERGY DETECTORS FOR FIVE YEARS IN SPACE}

\author{V. BECKMANN$^*$ on behalf of the INTEGRAL collaboration}

\address{INTEGRAL Science Data Centre, Ch. d'Ecogia 16,\\1290 Versoix, Switzerland\\
$^*$E-mail: Volker.Beckmann@obs.unige.ch\\
http://isdc.unige.ch/$\sim$beckmann}



\begin{abstract}
The {\it INTEGRAL\/} satellite, which studies the Universe in the hard
X-ray and soft Gamma-ray domain, has been operational for 5 years
now. The X-ray telescopes, which use the coded mask technique, provide
unprecedented spectral and imaging resolution.  
This led to a number of discoveries, such as the distribution of
diffuse emission in the Galaxy, the discovery of highly absorbed
sources and fast X-ray transients in the Galactic Plane, localization of $\sim50$ Gamma-ray
bursts, and the resolution of the cosmic X-ray background around its
peak at 30 keV. 
About 300 previously known X-ray sources have been detected and in
addition more than 200 new sources have been discovered. {\it INTEGRAL\/}
provides spectra starting at 3 keV and ranging up to several hundred keV. 
This article gives a brief overview about the major discoveries of {\it INTEGRAL}.
\end{abstract}


\bodymatter

\section{The INTEGRAL Mission}
ESA's {\it INTEGRAL\/} space mission \cite{INTEGRAL} hosts two major
hard X-ray instruments, IBIS and SPI, both coded-mask telescopes. IBIS
\cite{IBIS} provides imaging resolution of 12 arcmin, while SPI
\cite{SPI} is optimized for spectroscopy. Both instruments operate at
energies from 15 keV up to several MeV. Co-aligned with these main
instruments are the two X-ray monitors JEM-X \cite{JEMX}, which
provides spectra and images in the 3--30 keV band, and the
optical camera OMC \cite{OMC}, which provides photometry in the V
filter. {\it INTEGRAL} was launched on October 17, 2002 from Baikonur
into a highly eccentric orbit with a perigee of $9,000$ km and an
apogee of $150,000$ km, which avoids as much as possible the Earth's
radiation belt and allows for un-interrupted observations of up to 3
days. 

\section{Data processing for INTEGRAL and the ISDC}
Data from the {\it INTEGRAL\/} mission is made available to the
community via the {\it INTEGRAL Science Data Centre} (ISDC \cite{ISDC}). The
telemetry of the satellite is sent in a constant data flow of $\sim 90
\rm \, kbit \, s^{-1}$ to the Mission Operations Centre in Darmstadt. From there
the data are then sent to the ISDC in Versoix (near Geneva). Incoming data are searched for transient sources and
Gamma-Ray bursts within a few seconds after arrival by the {\it INTEGRAL
Burst Alert System} (IBAS \cite{IBAS}). The telemetry data are
pre-processed, which means that they are decoded and data are stored into
a fits file data structure. Afterwards the
data are analysed by a scientist on duty in order to look for
scientifically interesting events, such as the occurrence of new
sources, and to inform the astrophysical community. The data are then archived together with standard analysis
results and distributed to the scientific community. 
In addition, the ISDC provides scientific analysis software to the
community, in collaboration with the instrument teams.
The software, documentation and {\it INTEGRAL}'s data are available through
the ISDC web-page\footnote{The ISDC web page is located at
  http://isdc.unige.ch/}. 

\section{Scientific Highlights}

Within the first five years of the mission {\it INTEGRAL\/} related
research led to
several hundreds of scientific publications. Below I will present a
few of them, focussing on most recent discoveries. This list is obviously incomplete and biased, and I
refer the reader to the list of {\it INTEGRAL\/} related publications
maintained at ESA/ISOC\footnote{http://www.sciops.esa.int/index.php?project=INTEGRAL}.

\subsection{The Galactic Centre}

Using ISGRI \cite{ISGRI}, the soft $\gamma$-ray detector of IBIS, it was for
the first time possible to resolved the Galactic Centre above 20
keV. A new source, IGR~J1745.6--2901, was found to be the hard X-ray
counter part of Sgr~A*, which appears to be a faint
but persistent source at this energy with a spectrum which apparently originates from a two temperature plasma with $kT \simeq 1.0 \rm \, keV$ and $kT \simeq 6.6 \rm \, keV$ \cite{Belanger06}. The source seen by {\it INTEGRAL} is also detected at TeV energies by HESS \cite{SgrHESS}. ISGRI also
detected another new source close by, which has been associated with
the giant molecular cloud Sgr~B2 \cite{comptonmirror}. This cloud
might reflect the emission of Sgr~A* in the X-rays and therefore
function as a ``Compton mirror'' \cite{comptonmirror2}. If this hypothesis is true, the
emission of the central black hole of our Galaxy was much higher about
300 -- 400 years ago than it appears
today. 

\subsection{Compact Objects}

Galactic compact objects are the brightest persistent sources seen by
{\it INTEGRAL}. Therefore during the first five years of the mission,
the majority of observing time was used to cover the Galactic Centre
and the Galactic Plane, revealing not only a large number of new
sources, but also allowing for the first time to monitor in detail the
spectral evolution of known hard X-ray sources.
Within the class of black hole candidates especially Cygnus~X-1 led to
many new results, such as the detection of a hard tail in the spectrum
in excess of the thermal Comptonization \cite{CygX-1_Marion}, and the
complex nature of the physical processes involved in state transitions
\cite{CygX-1_Fritz,CygX-1_Malzac}. In the case of GRO~J1655--40, the
interpretation of a bright burst observed in 2005 is still ongoing, as
some studies find evidence for a spectral cut-off at high energies
\cite{Shaposhnikov07}, whereas others find an undisturbed power law up
to several hundreds of keV \cite{GRO1,GRO2}. In addition the number of
black hole candidates with well studied spectra increased
significantly with {\it INTEGRAL}. This includes the cases of XTE~J1550--564
where an underluminous outburst was observed \cite{Sturner05}, XTE J1817--330, in which the ISGRI data indicate a
thermal accretion disk and a comptonizing hot corona \cite{Sala07},
4U~1630-47 shows a variety of high-amplitude variability occurring
at the highest disk luminosities \cite{Tomsick05}, and also newly discovered black hole
candidates like IGR~J17464--3213 \cite{Capitanio05}.

Extending the spectra into the soft $\gamma$-ray
range, {\it INTEGRAL} shows that many low mass X-ray binaries (LMXB) indeed have
variable ``hard tails'' \cite{Farinelli07}, which most likely
originate in the Compton cloud
located inside the neutron star's magnetosphere \cite{Paizis06}. 
Concerning the pulsars, the enigmatic rotating neutron stars,
cyclotron lines can now be measured in greater detail, e.g. in
A~0535+26 \cite{Caballero07}, or in the anomalous X-ray Pulsar 4U~0142+614, which shows a complex spectrum and timing behaviour in the
hard X-rays \cite{Rea07}. The anomalous X-ray pulsars show very hard
and pulsed X-ray emission, which indicates that these sources are
indeed magnetars \cite{Kuiper06}.
The observation of the outburst of the pulsar V0332+53
\cite{Kreykenbohm05} has led to the discovery that the brightness
decline is accompanied by a change in the extent of the cyclotron
scattering region \cite{Mowlavi06}, gave detailed insight into its
orbital parameters \cite{Zhang05} and geometry \cite{Schoenherr07}, and revealed that the energies at
which the cyclotron
lines appear change linearly with the source luminosity \cite{Tsygankov06}.
Several pulsars have been shown to be accretion powered, such as the
newly discovered pulsars IGR~J00370+6122 \cite{intZand07} and
IGR~J18483-0311 \cite{Sguera07}, and 4U~1954+319 turned out to be a
symbiotic LMXB with the slowest wind-accreting X-ray pulsar ever
observed \cite{Mattana06}. On the other hand, {\it INTEGRAL} also
discovered the fastest millisecond pulsar ever known, the new source
IGR~J00291+5934 \cite{Shaw05}.
The accreting Be/X-ray pulsar SAX~J2103.5+4545 has been a puzzling object, as it was seen exhibiting hard/high and soft/low states \cite{Camero07} and large spin-up rate \cite{Sidoli05,Blay04}.

 
Microquasars have been also studied in great detail. GX~339--4 displays
a
variable high energy cut-off which might suggest that the
low- and high-energy components in this source have a different origin
\cite{Joinet07}. The X-ray binary LSI~+61~303 also seems to host a
microquasar, but does not reveal the presence of a cut-off, and the
observed spectrum and spectral variability can be explained if the
compact object in the system is a rotation powered pulsar
\cite{Masha06}. LS~5039, the only persistently detected X-ray binary at TeV energies, has been detected by IBIS/ISGRI and the data hint to a break in the spectral behaviour at hard X-rays \cite{Goldoni07}. 
For GRS~1915+105 new variability patterns have been
discovered \cite{Hannikainen05}, whereas the case of Cyg~X-3 and
whether or not it hosts a microquasar remains a question of debate
\cite{BeckmannCyg,Hjalmarsdotter07}.

\subsection{New Sources found by INTEGRAL}

Since the launch of the mission, {\it INTEGRAL\/} has discovered more
than 200 previously unknown hard X-ray sources\footnote{See the
  webpage listing the new sources discovered by {\it INTEGRAL\/} under\\
  http://isdc.unige.ch/$\sim$rodrigue/html/igrsources.html}. 
As
most of the new sources were discovered along the Galactic Plane,
optical identification of the counter part is often difficult. 
Several projects aim at this problem, for example by comparing the
{\it INTEGRAL} sources with photometric catalogues
\cite{Negueruela07} or with the ROSAT catalogues \cite{Stephen06},
by optical spectroscopy (e.g. \cite{Masetti06a,Masetti06b}), or by
simultaneous multi-telescope follow-up observations (e.g. \cite{BeckmannIGR}). Despite these efforts, more than 50\% of the new discoveries remain unclassified.
Among the Galactic sources, {\it INTEGRAL} discovered a number of
highly absorbed high mass X-ray binaries (HMXB), in which the binary system is obscured by the
strong stellar wind originating from the massive donor star
\cite{Walter06,Leyder07,BeckmannIGR}. Another study suggests that the obscured sources may be microquasars like SS~433, but with slightly lower mass transfer rate \cite{Begelman06}. Another class of rare objects found are HMXB with lower absorption, the Supergiant Fast X-ray Transients (SFXT). In these sources the X-ray emission might appear when the neutron star crosses the stellar wind of the giant star \cite{Chaty07}, or could be due to several wind components \cite{Sidoli07}. 
The majority of the new sources seems to be distributed in the Galactic Plane rather like the LMXB than the HMXB population \cite{Bodaghee07}.
The extended mission of {\it INTEGRAL} allows now also to detect
fainter sources, like the magnetic CV IGR~J00234+6141
of the intermediate polar type. The discovery of this type of object 
confirms
earlier conclusions that intermediate polars contribute significantly
to the population of galactic X-ray sources and represent a
significant fraction of the high energy background \cite{Bonnet-Bidaud}.
Lately it has also been shown that several unidentified sources
detected in the TeV range are counterparts of {\it INTEGRAL}
sources \cite{Ubertini05,Malizia05}. While for point-like and variable TeV sources the
correspondence with the {\it INTEGRAL} sources is almost sure, we seem
to observe different acceleration sites in the case of extended
sources like supernova remnants and pulsar wind nebula \cite{Walter07}.

In the extragalactic sky {\it INTEGRAL} detected more than 160 AGN, mainly
Seyfert galaxies, of which many were not known to be hard X-ray
emitters \cite{Bird07,Bodaghee07}. 

\subsection{Active Galactic Nuclei and the CXB}

With the wide spectral coverage, {\it INTEGRAL} was used to study in
detail several bright AGN. The AGN detected by IBIS/ISGRI above 20
keV are on average low
luminous and near-by ($\bar z = 0.02$) Seyfert galaxies
\cite{AGNcat,Bassanicat,Sazonov07}. In the hard X-rays the absorption by Galactic hydrogen does not affect the spectra, thus {\it INTEGRAL} can observe AGN shining through the Galactic Plane \cite{Soldi05,Molina06}. In most of the AGN it appears
that the spectral turn-over from a simple power law is located at
energies $\gg 100 \rm \, keV$, as shown in the Seyfert 2 galaxy NGC~4388 in which the
absorbing material is probably far from the central engine
\cite{NGC4388}, NGC~4151 \cite{NGC4151}, NGC 2992, the most variable
Seyfert of the bright AGN seen by {\it INTEGRAL} \cite{NGC2992}, Cen~A \cite{CenA}, and in a spectral
study of 72 AGN combining data from all three X-ray instruments on
board the satellite
\cite{AGNcat}. The objects with sufficient signal-to-noise
show though signatures of Compton reflection \cite{NGC4151,IGR21247}, but only a dozen AGN have been detected by IBIS/ISGRI above 100 keV so far \cite{Bazzano06}. Only about 10 blazars have been
detected and studied so far, such as S~0716+714 \cite{0716}, 3C~454.3
\cite{3C454}, which both were observed after an outburst. The lowest
redshift quasar 
3C~273 shows a historic minimum in its X-ray emission \cite{3C273,3C273b}, which allows to study the spectral features usually hidden by the strong jet emission \cite{3C273c}, and lately the farthest object seen so far by
{\it INTEGRAL} has been detected: IGR~J22517+2218 which is spatially coincident with
MG3~J225155+2217, a quasar at $z=3.668$ \cite{Bassani07}.

Related to the compilation of AGN surveys in the hard X-rays is the
question of what sources form the cosmic X-ray background (CXB) which
peaks at an energy of about $30 \rm \, keV$. 
An early study seemed to indicate that a large fraction of low
luminous AGN contributes to the CXB \cite{Krivonos05}, 
but apparently these source detections were spurious and could not be
confirmed by later studies. Number counts of AGN and the construction
of the X-ray luminosity function remains one of the major aspects of
current {\it INTEGRAL} related AGN research. The first luminosity function derived in this energy range indicated that the {\it INTEGRAL\/} detected AGN cannot explain the CXB without evolution of the AGN population \cite{Beckmannlumfunc}, and a recent study indicates that the CXB can be explained when considering luminosity dependent density evolution \cite{Sazonov07}. On the contrary, a study on the {\it XMM}-LSS field came to the conclusion that an evolution in absorption (towards stronger absorbed sources at higher redshifts) has to be assumed in order to connect the CXB to INTEGRAL detected AGN \cite{Treister07}.
In all these studies it has to be taken into account that the exact strength of the CXB is still under debate. {\it INTEGRAL\/} measurements of the CXB through Earth-occultation technique seem to indicate that the actual flux is higher by $\sim 20\%$ when compared to earlier HEAO-1 measurements \cite{Churazov07}.

\subsection{Gamma-Ray Bursts, SGRs, and the IBAS system} 

The {\it INTEGRAL\/} Burst Alert System (IBAS \cite{IBAS}) monitors the incoming data for events in the Anti-Coincidence System (ACS \cite{ACS}) of the spectrometer SPI and in the field of view of IBIS. For the latter events which have the signature of a Gamma-Ray Burst (GRB), the coordinates are determined automatically with high precision (usually $\sim 2$ arcmin uncertainty) and distributed within $\sim 20$ seconds to the astronomical community.  As of October 2007, {\it INTEGRAL} has detected 47 GRB in the field of view\footnote{see http://ibas.iasf-milano.inaf.it}. Among those GRB, only one event is a so-called short burst \cite{BeckmannGRB,Goetz07}. Because of the high sensitivity of IBIS/ISGRI, even faint GRBs are detected and thus {\it INTEGRAL} was able to see the sub-energetic GRB 031203, the closest to Earth determined so far \cite{Sazonov04,Soderberg04}. For the brightest one, GRB 041219a, it was possible to constrain the degree of linear polarisation $63 {+31 \atop -30} \%$ \cite{McGlynn07}. Thanks to IBAS, {\it INTEGRAL} was the first satellite in 2004 to report the giant flare from the soft gamma-ray repeater SGR~1806-20 \cite{Mereghetti05}. The bright flare of this magnetar, i.e. a strongly magnetized neutron star, can be explained by magnetic reconnection caused by crust breaking of the neutron star's surface \cite{Hurley05}. Up to now {\it INTEGRAL} has seen 4 confirmed SGRs and 3 promising candidates \cite{Diego06,GRB071017}.

The GRB detected by the SPI-ACS are also made public immediately through the web\footnote{For all alerts from the IBAS system see http://isdc.unige.ch/index.cgi?Soft+ibas}. Although no localization is available for these bursts, the ACS provides 50 ms lightcurves for on average one burst every 3 days \cite{Rau05}. 

\subsection{Diffuse Galactic Emission}

The Ge detectors of the SPI spectrometer \cite{SPI} provides a resolution of about 2.5 keV
at 1.3 MeV. This is sufficient to resolve Gamma-Ray lines of the
Galactic diffuse emission. A prominent line is the 511 keV
annihilation line of cosmic positrons. SPI measurements demonstrate
that the positrons mainly annihilate with free electrons and that the
medium must be at least partially ionized \cite{Jean06}. Another
finding is that most of the emission seems to arise from the Galactic bulge
and not from the Galactic disk \cite{Georg07}, which led to speculations that the
source of the diffuse emission is the Galactic black hole Sgr A*
\cite{Cheng07} or that it might origin from electroweak scale WIMPs
\cite{WIMPs}, from WIMP candidates with an ``excited state''
\cite{WIMPs2}, or from pulsar winds \cite{pulsarwinds}. 
Another bright Gamma-ray line at 1808.65 keV arises from the decay of radioactive
$^{26}$Al. Analysis of SPI data led to the finding that 
$^{26}$Al source regions corotate with the Galaxy, supporting its
Galaxy-wide origin and revealing a core collapse
supernova rate of $1.9 \pm 1.1$ per century in our Galaxy
\cite{Al26}. 
Other gamma-ray lines which have been observed from nucleosynthesis
sites include $^{44}$Ti \cite{Ti44} and $^{60}$Fe \cite{Fe60}.
Gamma-ray lines from point sources, as they were reported from
previous Gamma-ray missions with less sensitivity, have so far not
been detected and also no ``unexpected'' lines have been found
\cite{lines}.

Another mystery solved by {\it INTEGRAL\/} is the origin of the
Galactic background emission at 20--60 keV, which has been demonstrated
to arise from compact sources \cite{LebrunNature}, whereas the
Galactic ridge emission at soft $\gamma$-rays (200 keV -- 1 MeV) has
been shown to be indeed a diffuse or unresolved
emission component which might be caused by in-situ electron acceleration or by
an unresolved population of weak sources with hard X-ray spectra
\cite{ridge}.

\section{Outlook}

All instruments of the {\it INTEGRAL} mission are in good health. The recent results have shown that many science projects are only possible through the very deep exposures, available after several years of observation. Not only will the study of the diffuse emission benefit from an on-going {\it INTEGRAL} project, as it will reveal the true distribution of the positrons, $^{26}$Al, $^{60}$Fe and other elements resulting from nucleosynthesis in the Galaxy. Especially for source population studies, like for X-ray binaries like LMXB, HMXB, and CVs, and of course for AGN, the continuous observation of the sky with high spectral and spatial resolution above 20 keV will be crucial. Concerning the AGN the question what causes the CXB will be answered, and the true physics in the central engine of Seyfert galaxies and what role Compton reflection plays in there, will be revealed. 
For Galactic sources the observations of the X-ray binary population will help us to understand the formation of accretion disks around black holes, and their contribution to the Galactic Ridge emission. This will help to disentangle the source mix apparent in our Galaxy. Finally, obscured sources, both Galactic and extragalactic will be discovered, helping us to peer through the obscuring material into the true nature of the most violent physical processes.

\section*{Acknowledgments}

{\it INTEGRAL} is an ESA project funded
by ESA member states (especially the PI countries: Denmark, France,
Germany, Italy, Spain, Switzerland), Czech Republic, Poland, and with the
participation of Russia and the USA.


\end{document}